# Graphene-Ferroelectric Hybrid Structure for Flexible Transparent Electrodes

*Guang-Xin Ni[1,2,3], Yi Zheng[1,2], Sukang Bae[4,6], Chin Yaw Tan[5], Orhan Kahya[1], Jing Wu[1], Byung Hee Hong[4,6†], Kui Yao[5], and Barbaros Özyilmaz[1,2,3,7†]*

[1]Department of Physics, 2 Science Drive 3, National University of Singapore, Singapore 117542
[2]NanoCore, 4 Engineering Drive 3, National University of Singapore, Singapore 117576
[3]Graphene Research Centre, 6 Science Drive 2, National University of Singapore, Singapore 117546
[4]SKKU Advanced Institute of Nanotechnology (SAINT) and Center for Human Interface Nano Technology (HINT), Sungkyunkwan University, Suwon 440-746, Korea
[5]Institute of Material Research and Engineering (IMRE), A*STAR (Agency for Science, Technology and Research), 3 Research Link, Singapore 117602
[6]Department of Chemistry, Seoul National University, Seoul, 151-747, Korea
[7]NUS Graduate School for Integrative Sciences and Engineering (NGS), Singapore 117456

[†]Corresponding authors: barbaros@nus.edu.sg, byunghee@snu.ac.kr

ABSTRACT Graphene has exceptional optical, mechanical and electrical properties, making it an emerging material for novel optoelectronics, photonics and for flexible transparent electrode applications. However, the relatively high sheet resistance of graphene is a major constrain for many of these applications. Here we propose a new approach to achieve low sheet resistance in large-scale CVD monolayer graphene using non-volatile ferroelectric polymer gating. In this hybrid structure, large-scale graphene is heavily doped up to $3\times10^{13}$ cm$^{-2}$ by non-volatile ferroelectric dipoles, yielding a low sheet resistance of 120 Ω/□ at ambient conditions. The graphene-ferroelectric transparent conductors (GFeTCs) exhibit more than 95 % transmittance from the visible to the near infrared range owing to the highly transparent nature of the ferroelectric polymer. Together with its excellent mechanical flexibility, chemical inertness and the simple fabrication process of ferroelectric polymers, the proposed GFeTCs represent a new route towards large-scale graphene based transparent electrodes and optoelectronics.

Graphene keeps attracting much attention with enormous amount of experimental and theoretical activity, since its first micromechanical exfoliation in 2004.[1-4] As one atomic layer membrane, graphene is highly transparent (97.3 %) over a wide range of wavelengths from the visible to the near infrared (IR).[5] Owing to its covalent carbon-carbon bonding, graphene is also one of the strongest materials with a remarkably high Young's modulus of ~ 1 TPa.[6] The combination of its high transparency, wideband tunability and excellent mechanical properties make graphene a very promising candidate for flexible electronics, optoelectronics and phonotics.[7-9] The technical breakthrough of large-scale graphene synthesis has further accelerated the use of graphene films as transparent electrodes.[10,11]

To utilize graphene as transparent electrodes for applications such as solar cells[12], organic light emitting diodes,[13] touch panels and displays[14], the key challenge is to reduce the sheet resistance to values comparable with indium tin oxide (ITO), which provides the best known combination of transparency (> 90 %) and sheet resistance (< 100 $\Omega/\square$).[8,15] Conventional methods to reduce the sheet resistance like electrostatic doping of graphene requires complex fabrication steps of dielectric deposition and gate electrode preparations, which are not practical for doping large-scale graphene and consume power to maintain the doping levels.[12,14] Chemical doping has been shown to effectively reduce the sheet resistance of graphene.[16-19] However, the doping mechanism of chemical dopants is not yet fully understood and the relationship between charge density and carrier mobility is still under debate.[20-22] Furthermore, the adsorption of moisture and other chemical molecules after chemical treatment leads to a 40 % increase of the graphene sheet resistance within few days.[23,24] Consequently, an additional carefully chosen thin polymer coating is necessary to maintain its high conductivity without compromising its high transparency.[24] Therefore, new approaches with improved performance, zero power consumption and simplified fabrication processes are highly desired.

Ferroelectric polymers such as poly(vinylidene fluoride-co-trifluoroethylene) (P(VDF-TrFE)) have been extensively explored for ultrafast, high density, non-volatile memories.[25-27] Recently, its potential has also been recognized for other applications, such as high efficiency organic solar cells,[28] high power energy storage[29] and protective coating layer.[30] The key to many ferroelectric applications is, that ferroelectric polymers can introduce a large non-volatile doping. Also important to note is, that they are chemically inert under ambient conditions. Taking graphene-ferroelectric non-volatile memory as an example, the non-destructive and non-volatile electrostatic doping of graphene is only limited by the remnant polarization of P(VDF-TrFE) (8 µC/cm$^2$), which is equivalent to a charge carrier density of $5\times10^{13}$/cm$^2$. Equally important, P(VDF-TrFE) thin films are essentially transparent (> 98 %) across the visible spectrum.[31] In addition, with appropriate thickness the fully polarized P(VDF-TrFE) thin films simultaneously do both, dope and provide excellent mechanical support to graphene. With the exception of Ref. [32], all previous approaches of chemically doping graphene had to be combined with a polymer supporting layer in a separate process step.[16,23] Thus, the combination of P(VDF-TrFE) with large-scale graphene provides an ideal solution for graphene for transparent conductor applications. However, to date no experimental studies have been performed along this direction.

In this paper, we present a new route, which achieves low sheet resistance values in large-scale single layer graphene by introducing a transparent thin ferroelectric (P(VDF-TrFE)) polymer coating. Such graphene-ferroelectric transparent conductors (GFeTCs) exhibit a low sheet resistance value down to 120 Ω/□ at ambient conditions due to a large electrostatic *non-volatile doping* of up to $3\times10^{13}$/cm$^2$ from ferroelectric dipoles. Beyond having low sheet resistance values, the GFeTCs are also highly transparent (> 95 %) in the visible wavelength range, making them suitable for optoelectronics applications where a combination of both is required. With the excellent mechanical support of P(VDF-TrFE), the hybrid GFeTCs fabrication can also easily be integrated with industrial scale fabrication processes such as roll-to-roll techniques. Furthermore, the limiting factors to achieve even lower sheet resistances in large-scale graphene are analyzed by means of low temperature measurements. Our results show that, once CVD graphene synthesis and transfer processes are optimized even lower sheet resistances are feasible without degrading the optical transparency.

RESULTS AND DISCUSSION

The device fabrication procedures of GFeTCs for sheet resistance ($R_S$) and optical transparency (T) measurements are illustrated in Fig. 1 (Details see Methods). For $R_S$ characterization, the device fabrication begins by patterning large scale devices on Si/SiO$_2$ substrate. After coating the P(VDF-TrFE) thin film on top of graphene, field effect transistor devices are formed by contacting them with a top gate. Their $R_S$ as a function of P(VDF-TrFE) polarization is ready to be characterized, and the corresponding device structure is illustrated in Fig. 1(iv). For T measurements and subsequent transport measurements on such transparent substrates the corresponding device diagram is shown in Fig. 1(vii). Note that graphene-P(VDF-TrFE) hybrid structures can also be free-standing due to the excellent mechanical support from P(VDF-TrFE) thin films, as shown in the same figure. For large scale application corona poling can be used for simple, contact free and large-scale polarization.[33] The latter can be easily integrated with a roll-to-roll process.[16]

Fig. 2a shows a wafer-scale array of large graphene field effect transistor devices on a 500 nm Si/SiO$_2$ substrate. In each unit cell, the graphene area is 1.44 mm$^2$, which is 10$^6$ times larger than typical CVD graphene devices (~ 3 μm$^2$).[34,35] The mobility (μ) of such large scale devices at room temperature varies between 2000-4000 cm$^2$/Vs. Here we discuss two representative samples in detail with a mobility of ~ 2000 cm$^2$/Vs before spin coating of P(VDF-TrFE) (see supplementary information). The coating of the P(VDF-TrFE) dielectric is followed by the definition of top contacts; the final device structure in the van der Pauw measurement geometry is shown Fig. 2b. The mobility of both devices remains ~ 2000 cm$^2$/Vs, *i.e.* comparable to the initial values. Fig. 2c shows the typical hysteresis polarization loops of P(VDF-TrFE) thin films as a function of the applied electric field. The key parameters relating to the value of sheet resistance are the maximum polarization ($P_{max}$) and remnant polarization ($P_r$), which are recorded as a function of the electric field (Fig. 2d). We start our discussion with the hole doping case. Both -$P_{max}$ and -$P_r$ increase and finally saturate with increasing applied electric field. The electrostatic doping level in graphene is n($V_{P(VDF-TrFE)}$) = β$P_r$/e (and n($V_{P(VDF-TrFE)}$) = β$P_{max}$/e) for -$P_r$ (and -$P_{max}$), where β is the electrical coupling between ferroelectric dipoles and graphene.[26] Fig. 2e and Fig. 2f show how $R_S$ varies as a function of increasing |-$P_{max}$| and |-$P_r$|, respectively. Prior to full polarization the sheet resistance of such samples is rather high ($R_S$ = 1440 Ω/□) due to the large disorder created by randomly oriented dipoles. A twelve fold reduction of $R_S$ is only achieved when P(VDF-TrFE) is fully polarized, resulting in a low sheet resistance of 120 Ω/□ for a single layer of graphene (Fig. 2f).[36] Even lower sheet resistances could be achieved at -$P_{max}$. However, this is of little practical value, since a constant voltage needs to be applied. The key advantage of P(VDF-TrFE) is indeed that after it is fully polarized, the induced non-volatile doping allows the sheet resistance of large-scale graphene to remain low, even when the power is turned off (SFig. 3 of Supplementary Information).

Besides low sheet resistance, high optical transparency and mechanical flexibility are equally critical for the widespread application of graphene based transparent electrodes in optoelectronics. The graphene-P(VDF-TrFE) hybrid structure for optical and electromechanical measurements are shown in Fig. 3. Fig. 3a shows the GFeTCs on top of PET substrates. The P(VDF-TrFE) film used here is only 1 μm thick, and yet already sufficient to provide an excellent mechanical support for graphene (Inset of Fig. 3a). The transmission spectra as a function of wavelength from the visible to near IR are shown in Fig. 3b. For a P(VDF-TrFE) thin film alone without graphene, interference effects lead to an oscillating transmittance feature.[37] Such interference effects imply a uniform P(VDF-TrFE) thin film. From the periodicity, the thickness of P(VDF-TrFE) thin film can also be deduced to be around 1 μm. This is in good agreement with independent surface profile measurements. On the other hand, once graphene is transferred on P(VDF-TrFE) films, the interference effect vanishes and a monotonic transmittance as a function of wavelength is observed. In the visible range the optical transparency of graphene-P(VDF-TrFE) hybrid structure ranges from 92.5 % to 96.3 % with 95 % at 550 nm.

Next we discuss the mechanical properties of GFeTCs. Fig. 3c shows the evaluation of GFeTCs foldability on PET substrate by measuring the resistance of graphene with respect to bending radius. The resistance showed a small increase down to the bending radius of 3.0 mm, which was recovered completely after unbending the GFeTC device. Notably, the original resistance is repeatedly restored even after bending the sample down to 1.0 mm in radius (approximate tensile strain of 11 %). These outstanding mechanical properties of graphene-P(VDF-TrFE) films are comparable with the results obtained for graphene films on PET.[10]

Temperature dependent sheet resistance measurements of large-scale CVD graphene at different doping levels were carried out to investigate the factors preventing even lower $R_S$. A clear transition from insulating behaviour at low doping (before polarizing P(VDF-TrFE)) to metallic behaviour at high doping (after fully polarizing P(VDF-TrFE)) was observed (Fig 4a). For samples of comparable mobility, previously such a behaviour has been associated with a large inhomogeniety of CVD graphene specific charged impurities.[38,39] More relevant for our purpose are temperature dependent resistivity measurements at high doping levels; At 50 K, once the phonon contribution has been eliminated ($\Delta R$ = 30 Ω/□), a residual sheet resistance of 100 Ω/□ is observed. This clearly shows, that at high doping much lower sheet resistances can be achieved once CVD graphene specific charged impurities mainly due to the etching and transfer processes are reduced or eliminated.[40] Note also, that high doping experiments with exfoliated graphene samples allow sheet resistances as low as 30 Ω/□ at 50 K.[41]

We can better understand our results by calculating the sheet resistance as a function of carrier density with $R_S = R_0 + R_{AP} + R_{FP}$. Here $R_0$ represents the residual sheet resistance due to extrinsic scattering sources,[42,43] the acoustic phonon scattering $R_{AP}$ gives rise to a linear T dependent resistivity[44] and $R_{FP} \sim T^2/(n/10^{12} + \gamma)$ represents the contributions from flexural phonons.[45] Using this formula, $R_S$ *vs* n can be plotted for mobilities ranging from 2,000 to 10,000 cm$^2$/Vs. With this we can now compare the $R_S$ values obtained at different n for two different samples (Fig. 4b). For both samples the experimental data can be well explained, if one assumes $\mu \sim 2{,}000$ cm$^2$/Vs. Since this value is comparable to the device mobility before the spin-coating of the ferroelectric polymer (~2,300 cm$^2$/Vs), we conclude that the latter is currently not limited by the small content of non-ferroelectric phase in P(VDF-TrFE).[26] Last but not least, we further estimate $R_S$ for high mobility samples at room temperature. As expected the sheet resistance decreases with increasing mobility and can reach for realistic device mobilities of 10,000 cm$^2$/Vs low values of 50 $\Omega/\square$ at $3 \times 10^{13}$ cm$^{-2}$. Note also, that at least at room temperature high doping may avoid the need for combining graphene with BN; even a doubling of the device mobility to 20,000 cm$^2$/Vs would reduce only marginally the sheet resistance any further.[46,47]

The presence of a large inhomogenous background doping in large-scale GFeTCs becomes even more evident in $R_S$ *vs* +$P_r$ measurements (Fig. 5a). Compared to a monotonic decrease of $R_S$ with -$P_r$, a continuous increase of $R_S$ with increasing +$P_r$ is observed instead. Note that also the corresponding $\mu$ differs strongly at -$P_r$ & +$P_r$, as shown in Fig. 5b. Such a strong asymmetry in both $R_S$ and $\mu$ is only observed in mm$^2$ size large-scale samples and absent in our µm$^2$ size devices. This implies the existence of a non-uniform p-type background doping in large-scale CVD graphene after transfer. A simple model sketched in Fig. 5c&d illustrates this scenario. When ferroelectric dipoles are tuned to be +$P_r$, the corresponding electrostatic n type doping in graphene gives rise to a random array of p-n junctions resulting in large potential steps along the current path (Fig. 5c). As a consequence, sheet resistance is strongly enhanced. On the other hand, for -$P_r$ we obtain p-p' junctions (Fig. 5d) leading to a much smoother potential landscape and hence, a much lower sheet resistance.

For large-scale applications, the lamination of CVD graphene with commercially available pre-polarized P(VDF-TrFE) foils is more intriguing. Utilizing such foils, low sheet resistances of large-scale graphene can be immediately achieved without further ferroelectric poling. This can largely simplify the device fabrication steps. Fig. 6a shows the XRD measurements of both pre-polarized P(VDF-TrFE) foil and in-house-produced P(VDF-TrFE) thin film. Both show a pronounced diffraction peak which clearly indicates that the ferroelectric phase (β-phase) is highly crystalline.[48] The doping induced on graphene by the un-polarized and pre-polarized P(VDF-TrFE) foils is further determined by the Raman spectroscopy measurements (Fig. 6b). For graphene on pre-polarized P(VDF-TrFE) foil, a G peak shift of around 12 cm$^{-1}$ is observed, corresponding to ~ $1\times10^{13}$ cm$^{-1}$ electron doping.[49] At this doping level, the I-V measurements yields a sheet resistance value of ~ 500 Ω/□ (Fig. 6c). This value, while ~ 5 times larger than what has been achieved with in-house-produced P(VDF-TrFE) is promising, since the results can be further improved by optimizing the remnant polarization of commercial P(VDF-TrFE) sheets, the graphene on P(VDF-TrFE) transfer process and by graphene/P(VDF-TrFE) interface engineering (See Supplementary Information). Equally important, both n-type and p-type doped graphene films can be realized by transferring them to either side of pre-polarized P(VDF-TrFE) foils. Such graphene films can, *e.g.* in principle directly serve as anode and cathode in photovoltaic devices with controllable polarity.[26] For instance, the realization of n-type doped graphene is attractive for low work function cathode in light emitting diodes and solar cell devices.[50]

CONCLUSIONS

In conclusion, we have demonstrated a new type of transparent conductor using graphene-ferroelectric hybrid films. The ferroelectric thin film does not compromise the high optical transparency of graphene. It provides *non-volatile* electrostatic doping, yielding even in low mobility samples a sheet resistance as low as 120 Ω/□. The ferroelectric polymer also serves as an excellent mechanical supporting layer, making GFeTCs easily transferable and integrable with flexible electronics, optoelectronics and photonics platforms. In addition, we show that the limiting factors for further lowering the sheet resistance are not the ferroelectric polymer but commonly known charged impurities originating from existing transfer processes. Therefore, with further improvements in the transfer process a sheet resistance of 50 Ω/□ at optical transparency >95 % seems feasible.

METHODS

**Device Fabrications.** Device fabrication begins with the large-scale graphene synthesized by the CVD method on pure copper foils (Fig. 1 (i)), the details of graphene fabrication procedures are discussed in Ref [11, 16]. For the sheet resistance measurements, GFeTC devices are fabricated on conventional Si/SiO$_2$ (500nm and 300 nm) substrates as shown from Fig. 1 (ii) to (iv). (ii) The Cu-CVD graphene was immersed into copper etchant. (iii) Once the copper was etched away, the large-scale graphene was immediately transferred onto the substrates followed by standard e-beam lithography (EBL) and oxygen plasma etching process to separate large-scale graphene sheets into 1.2 mm by 1.2 mm graphene squares. After this, metal contacts (5 nm Cr/30 nm Au) were defined using the pre-defined shielding mask followed by thermal evaporation process. The devices were further thermally annealed at 250°C in H$_2$/Ar conditions for 3 hours. (iv) After spin coating 1.0 μm thick poly(vinylidene fluoride-trifluoroethylene) (P(VDF-TrFE)) followed by thermally evaporating the top gate electrodes, samples were ready to be characterized.

Fig. 1 (v) to (vii) summarises the GFeTC devices for the optical transparency measurements. (v) P(VDF-TrFE) thin film was spin-coated directly on top of graphene, which serves as the dielectric as well as mechanical supporting layer in the following steps. (vi) The Cu-CVD graphene-P(VDF-TrFE) structure was immersed into copper etchant. (vii) After removing the copper and rinsingin DI water, the graphene-P(VDF-TrFE) hybrid structure was immediately transferred to transparent substrates, *i.e.*, PET or glass substrates. For the optical measurements, the transparency of graphene-P(VDF-TrFE) *versus* wavelength is characterized using UV Probe 3600 at ambient conditions. The incident power intensity is 5000 mW.

**Transport Measurements.** Transport measurements were electrically characterized in ambient conditions with standard van der Pauw configuration using a lock-in amplifier with an excitation current of 100 nA. In total 6 single-layer large-scale graphene devices have been measured. Here we discuss 2 representative SLG devices in more details. For the temperature dependent measurements, samples were characterized in a variable temperature insert in a liquid Helium cryostat (T = 2-300 K).

To polarize the ferroelectric thin film, we mainly utilized the ferroelectric Radiant polarizer which injects a voltage pulse either positive or negative to the top gate electrode. After this, the corresponding resistivity of large-scale graphene at each $P_r$ magnitude of ferroelectric thin film was recorded using the van der Pauw approach, as shown in Fig. 2f. Note that for the van der Pauw measurements, the horizontal direction resisvitities and vertical direction resistivities were first compared to make sure their differences are negligible. Then the sheet resistance is calculated using the standard formula ($R_S = \pi R/\ln 2 \approx 4.53R$).

**Optical Measurements.** The linear optical transparency was measured using the UV-Vis-NIR spectrophotometer. The spectra are collected in a 1.0 mm path length cell. To perform these measurements, the graphene/ferroelectric hybrid structures are transferred to glass substrate.

**Raman Spectroscopy.** Raman spectroscopy/imaging were carried out with a WITEC CRM200 Raman system with 532 nm (2.33 eV) excitation and laser power at sample below 0.1 mW to avoid laser induced heating. A 50× objective lens with a NA=0.95 was used in the Raman experiments. Data analysis was done using WITec Project software.

**XRD Measurements.** The model of the XRD setup is Bruker Discover D8; The X-ray source: Cu-K(alpha) line (lambda = 1.541838 Angstrom).

*Acknowledgements.* The authors gratefully acknowledge useful help with Natarajan Chandrasekhar, Cedric Troadec, Yuda Ho, Chee-Tat Toh and Amar Srivastava for their help for the device preparations. This work is supported by the Singapore National Research Foundation grants NRF-RF2008-07, and NRF-CRP(R-144-000-295-281), NRF-POC002-038, NUS-YIA(R144-000-283-101), IMRE/10-1C0109, NUS/SMF grant, US Office of Naval Research (ONR and ONR Global), A*STAR SERC TSRP-Integrated Nano-photo-Bio Interface (R-144-000-275-305), NUS NanoCore, the National Research Foundation of Korea (NRF) funded by the Ministry of Education, Science and Technology (Global Research Lab. 20110021972, 2011K000615, 20110017587, 20110006268, and Global Frontier Research Program 20110031629).

*Supporting Information Available:* The contents of Supporting Information include the following: (1) Electrical transport characterization of large-scale CVD graphene. (2) Graphene device fabrication on commercially available pre-polarized P(VDF-TrFE) sheets. (3) Retention characterization. This information is available free of charge *via* the Internet at http://pubs.acs.org.

FIGURE CAPTIONS

**Fig. 1**. Schematic illustration of the GFeTC devices fabrication process for both sheet resistance characterization and optical transparency measurements (see text for details). (i) shows the large-scale CVD graphene on copper substrate, which is the starting point of all devices. From (ii) to (iv), device fabrications for $R_S$ measurements are illustrated. From (v) to (vii), device fabrication of transferring graphene-P(VDF-TrFE) hybrid structure on transparent substrate for transmittance measurements.

**Fig. 2**. Devices and measurement results of $R_S$ at ambient conditions. (a) Optical image of large-scale graphene samples with bottom contacts, the scale bar is 500 μm; (b) Optical image of a completed GFeTC device and its corresponding characterization strategy using van der Pauw method; (c) Typical hysteresis polarization loops of our P(VDF-TrFE) ferroelectric dielectric; The solid ball in blue, green and red colour represent low, medium and high level of remnant polarization ($-P_r$), respectively; The hollow ball in blue, green and red colour represent low, medium and high level of spontaneous polarization ($-P_{max}$), respectively; (d) Plot of $-P_r$ and $-P_{max}$ as a function of external electric field; The blue, red and green colour corresponding to the three different remnant polarization states. (e) Systematic gate sweep of $R_S$ as a function of $-P_{max}$. (f) $R_S$ as a function of $-P_r$.

**Fig. 3**. (a) Optical image of GFeTC samples on the transparent PET substrate; Inset shows the optical image of a free-standing graphene-P(VDF-TrFE) film hold by tweezers, the background is the logo of National University of Singapore (NUS); (b) Light transmission through graphene-P(VDF-TrFE) hybrid structure and pure P(VDF-TrFE) thin film as a function of wavelength from visible to UV regime. The red curve shows the optical image of pure P(VDF-TrFE) thin film. The vibration of P(VDF-TrFE) is due to the interference effect. (c) Mechanical foldability measurement of GFeTCs on PET (200 μm) substrate. Inset shows the optical image of the four-probe bending measurement setup.

**Fig. 4.** (a) Temperature-dependence measurements of GFeTCs at different charge density levels; the blue solid curve indicate the insulating behaviour of GFeTCs at low density level; the red solid curve indicate the metallic behaviour of GFeTC at high density level. (b) Experimental data and theoretical estimations of $R_S$ as a function of μ and n.

**Fig. 5**. (a) $R_S$ as a function of both $+P_r$ and $-P_r$. (b) The corresponding conductivity as a function of both $+P_r$ and $-P_r$. (c) Electrostatic doping in graphene with P(VDF-TrFE) at $+P_r$ state. The green colour particles in graphene represent the initial p-type charged impurities doping. After fully polarize the ferroelectric thin film, the formation of n-p electron-hole puddles in graphene explains the results observed in (a) and (b). (d) Electrostatic doping in graphene with P(VDF-TrFE) at $-P_r$ state.

**Fig. 6.** (a) XRD results of both P(VDF-TrFE) foil and house-produced P(VDF-TrFE) film. Inset shows the optical image of large-scale pre-polarized P(VDF-TrFE) foil on PET substrate. (b) Raman spectra of graphene on un-polarized (black colour) and pre-polarized (red colour) P(VDF-TrFE) foils. For clarity, we focus on the graphene G peak shift from 1580 cm$^{-1}$ (un-polarized) to 1592 cm$^{-1}$ (pre-polarized) as a comparison of the doping effect from P(VDF-TrFE) foil. (c) I-V characterization of the sheet resistance

of large-scale graphene on the pre-polarized P(VDF-TrFE) foil. Inset shows the device characterization image.

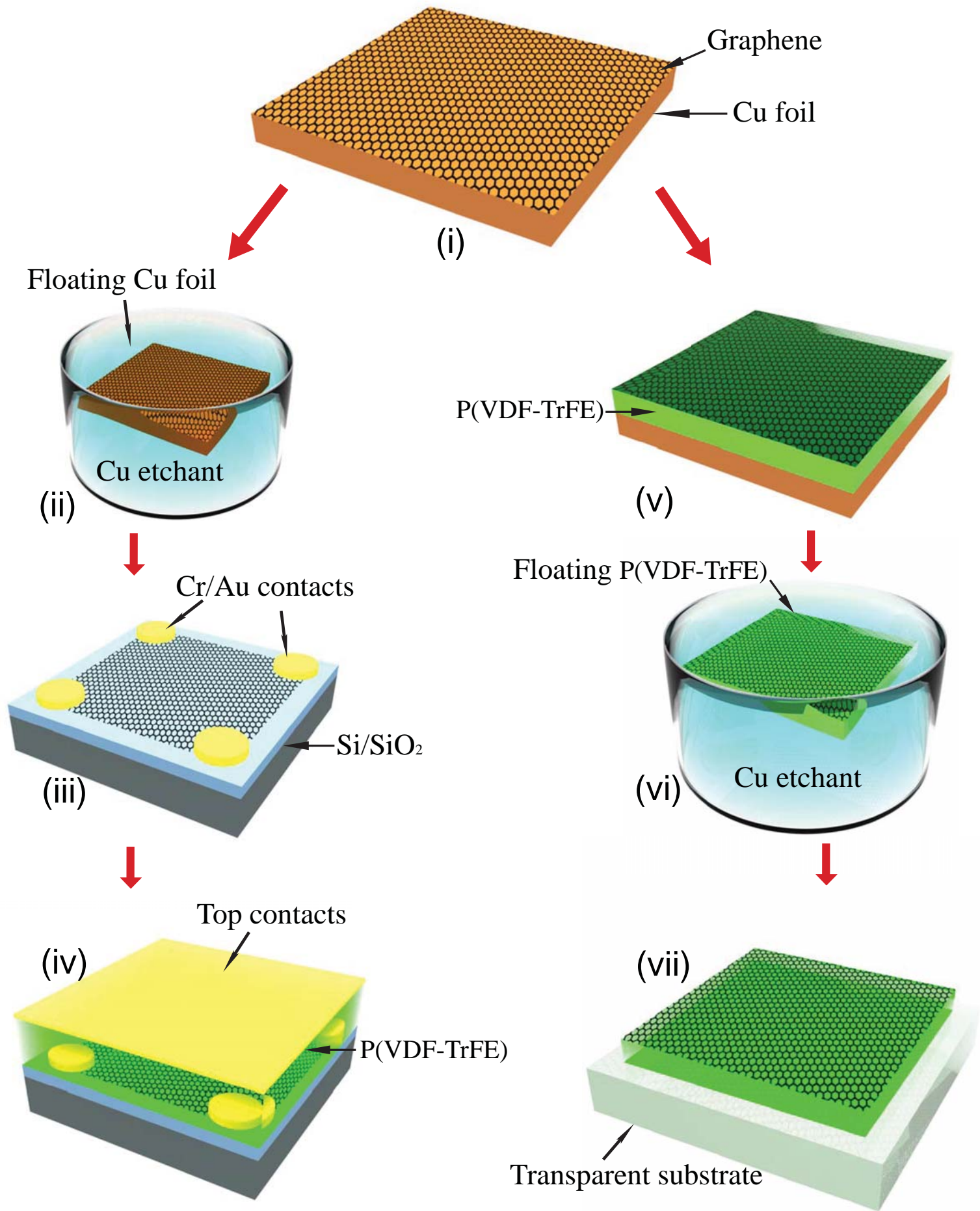

Fig.1 Guangxin Ni *et al*.

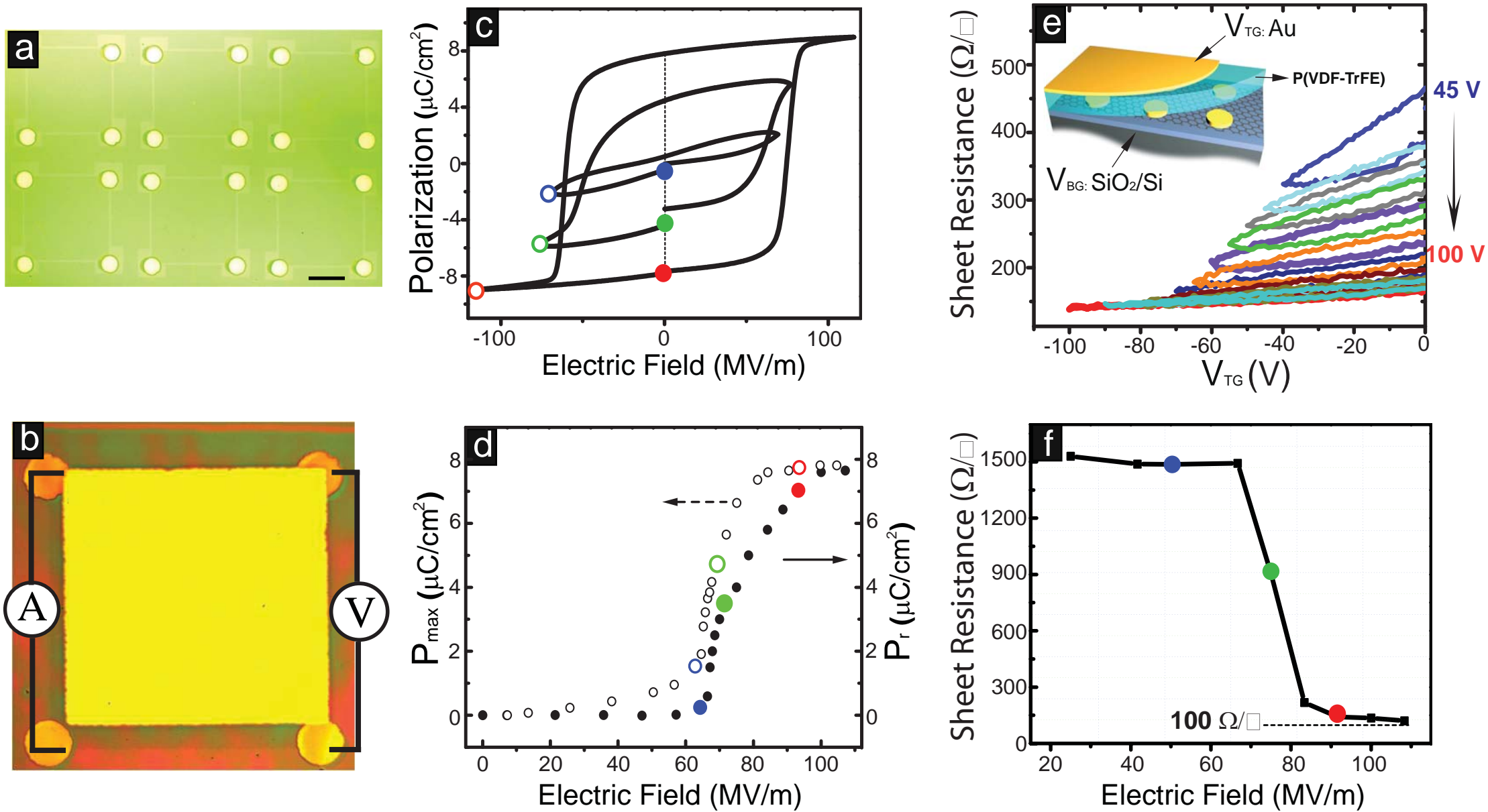

Fig.2 Guangxin Ni *et al.*

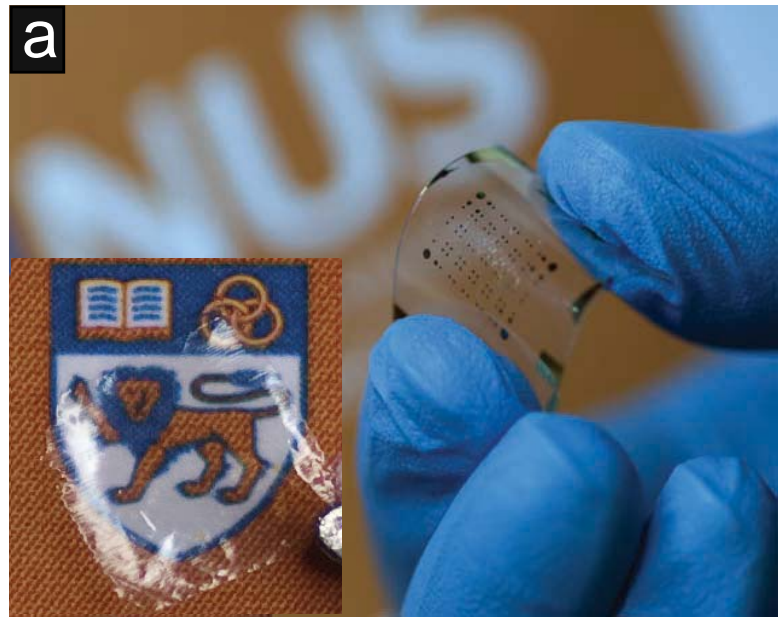 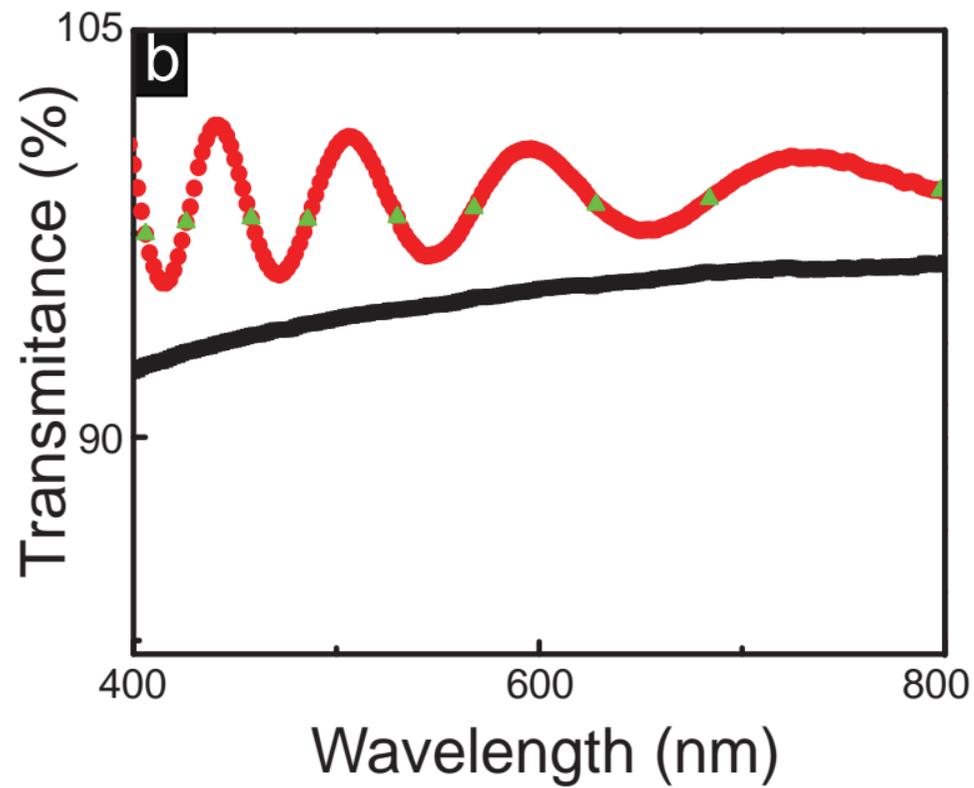 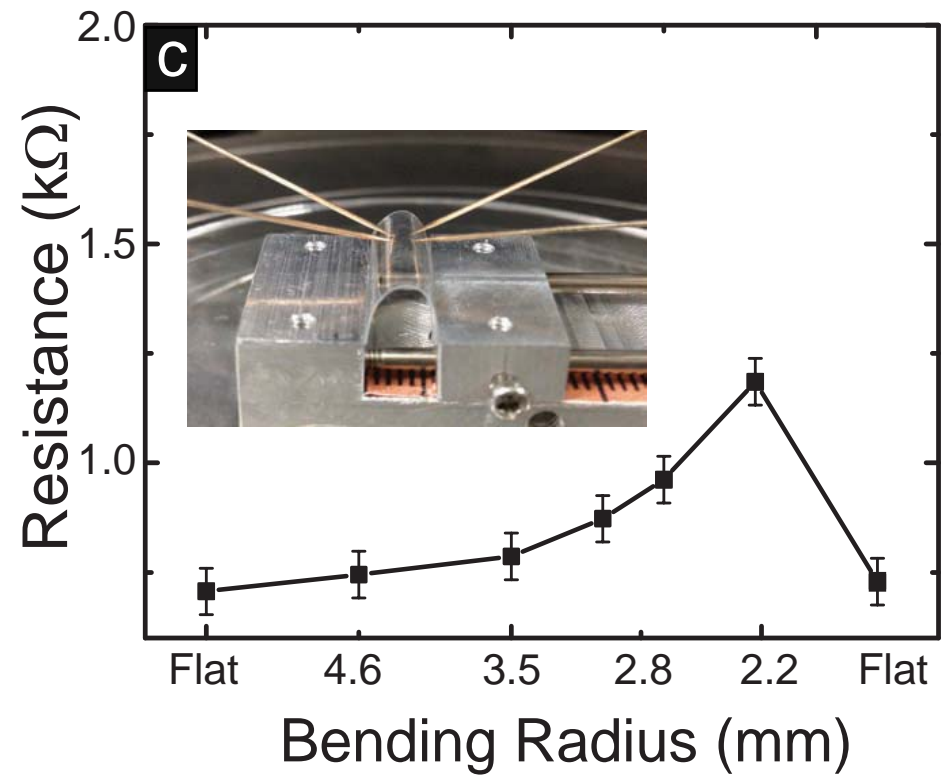

Fig.3 Guangxin Ni *et al*.

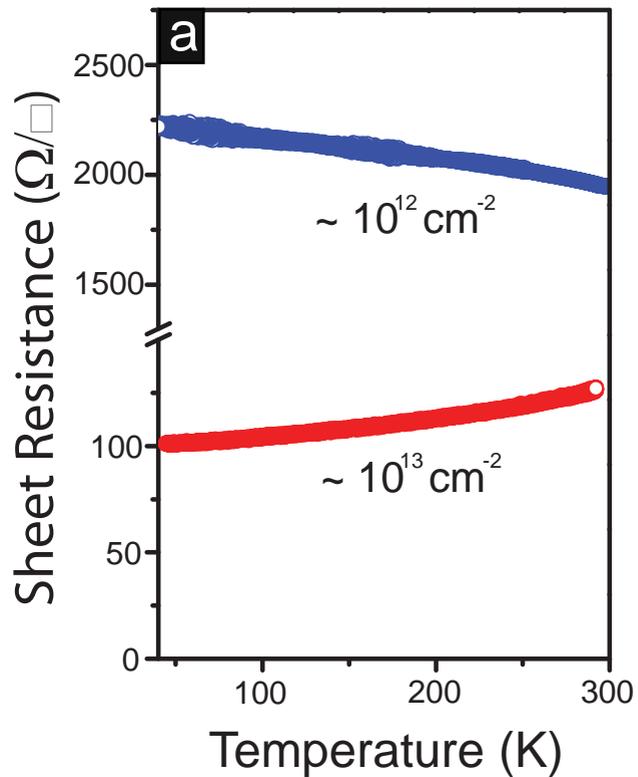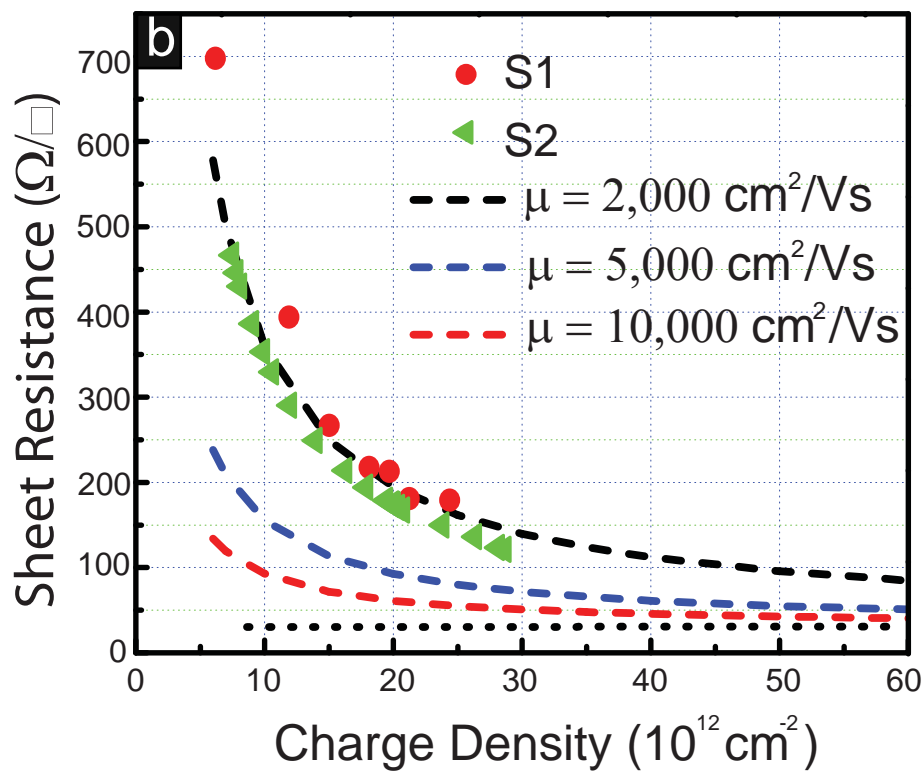

Fig.4 Guangxin Ni *et al.*

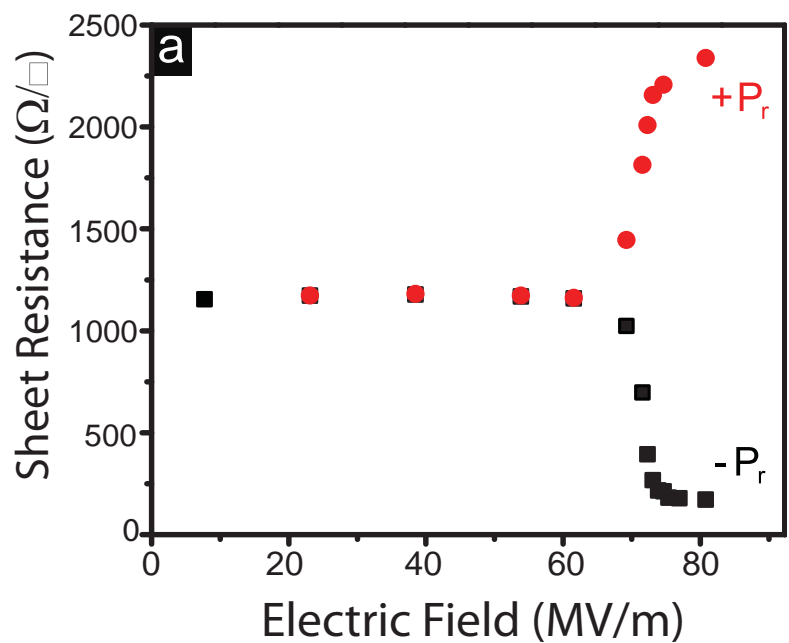
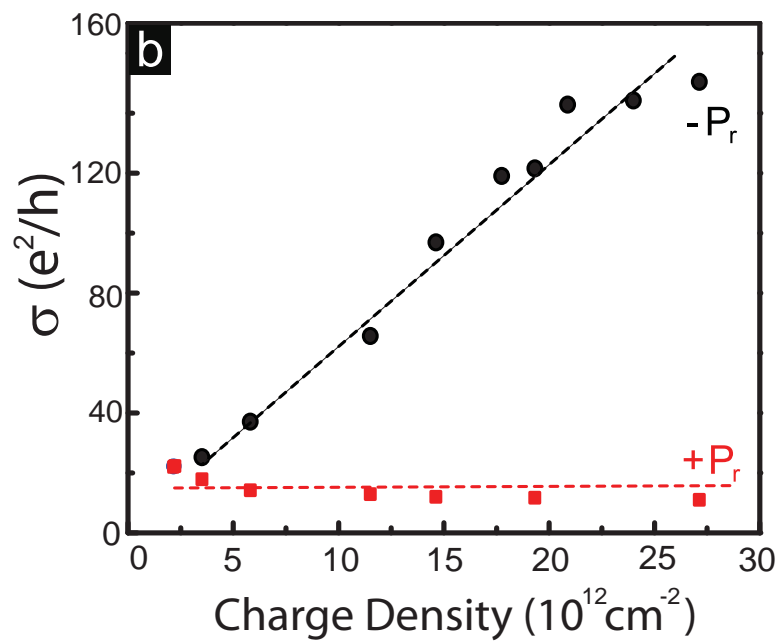
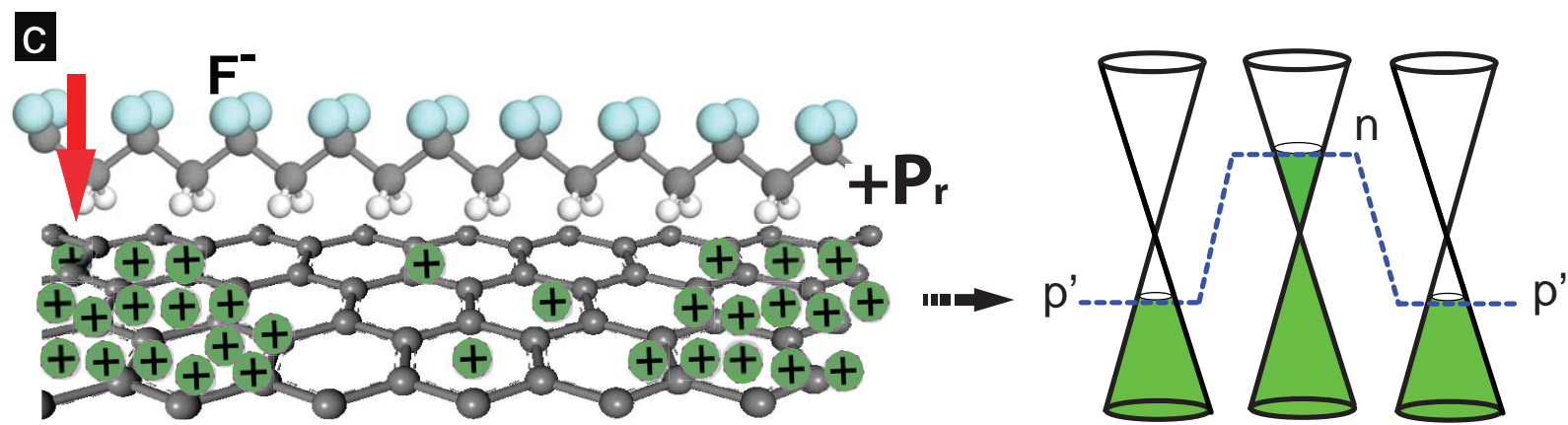
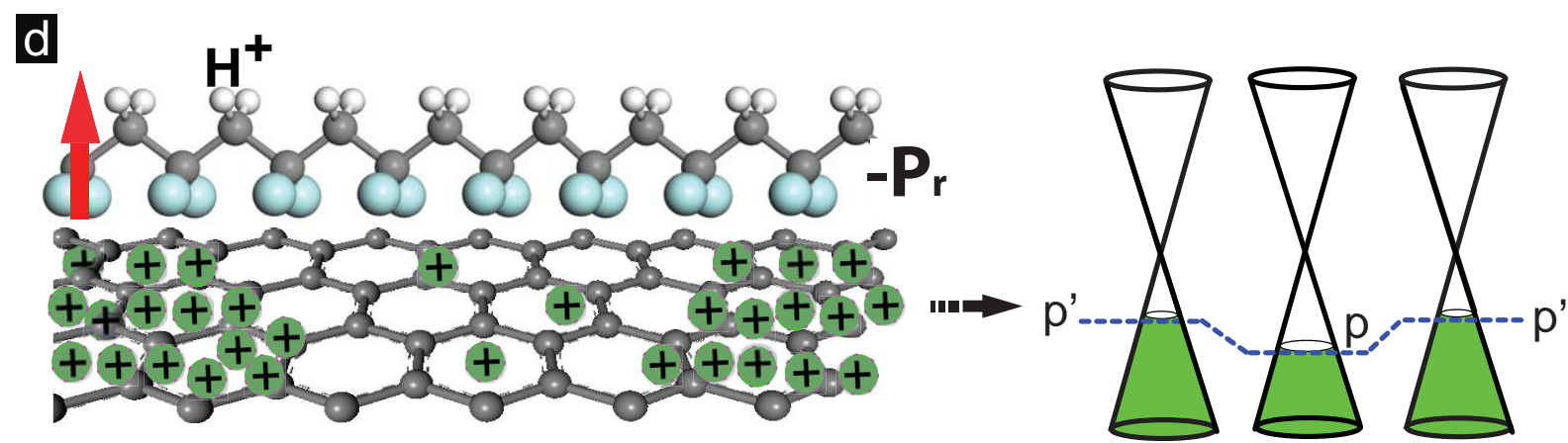

Fig. 5 Guangxin Ni *et al.*

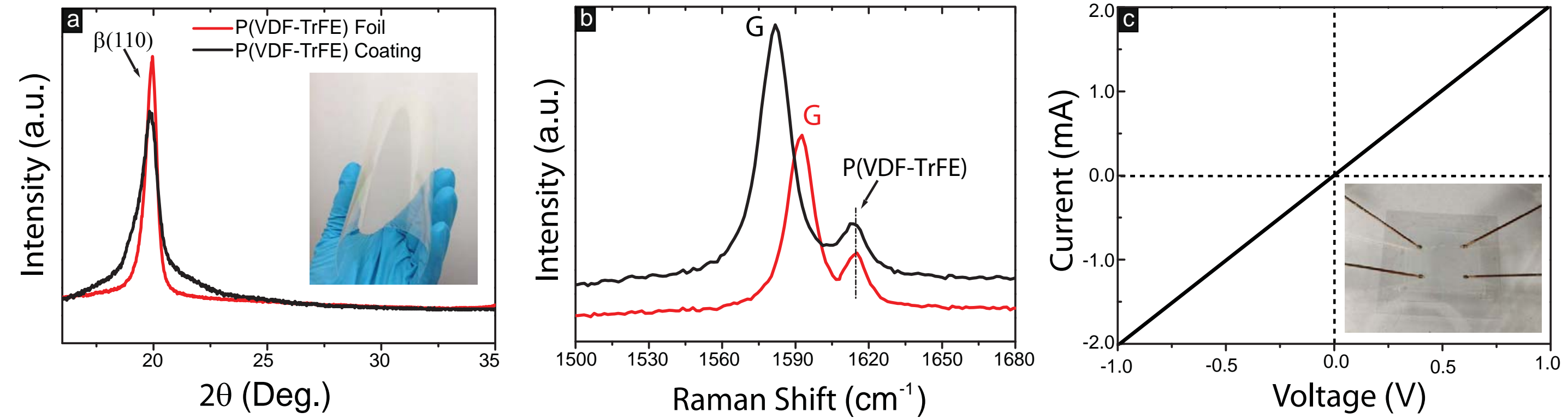

Fig.6 Guangxin Ni *et al*.